# Parallel Network with Channel Attention and Post-Processing for Carotid Arteries Vulnerable Plaque Segmentation in Ultrasound Images

Yanchao Yuan[1,2,3], Cancheng Li[1,2,3], Lu Xu[1,2,3], Ke Zhang[4], Yang Hua*[4,5,6], Jicong Zhang*[1,2,3]

1 School of Biological Science and Medical Engineering, Beihang University, Beijing, China
2 Hefei Innovation Research Institute, Beihang University, Hefei, China
3 Beijing Advanced Innovation Centre for Biomedical Engineering, Beihang University, Beijing, China
4 Department of Vascular Ultrasonography, XuanWu Hospital, Capital Medical University, Beijing, China
5 Beijing Diagnostic Center of Vascular Ultrasound, Beijing, China
6 Center of Vascular Ultrasonography, Beijing Institute of Brain Disorders, Collaborative Innovation Center for Brain Disorders, Capital Medical University, Beijing, China

**Abstract**

Carotid arteries vulnerable plaques are a crucial factor in the screening of atherosclerosis by ultrasound technique. However, the plaques are contaminated by various noises such as artifact, speckle noise, and manual segmentation may be time-consuming. This paper proposes an automatic convolutional neural network (CNN) method for plaque segmentation in carotid ultrasound images using a small dataset. First, a parallel network with three independent scale decoders is utilized as our base segmentation network, pyramid dilation convolutions are used to enlarge receptive fields in the three segmentation sub-networks. Subsequently, the three decoders are merged to be rectified in channels by SENet. Thirdly, in test stage, the initially segmented plaque is refined by the max contour morphology post-processing to obtain the final plaque. Moreover, three loss function Dice loss, SSIM loss and cross-entropy loss are compared to segment plaques. Test results show that the proposed method with dice loss function yields a Dice value of 0.820, an IoU of 0.701, Acc of 0.969, and modified Hausdorff distance (MHD) of 1.43 for 30 vulnerable cases of plaques, it outperforms some of the conventional CNN-based methods on these metrics. Additionally, we apply an ablation experiment to show the validity of each proposed module. Our study provides some reference for similar researches and may be useful in actual applications for plaque segmentation of ultrasound carotid arteries.

**Index Terms**: Carotid artery ultrasound, CNN, Morphology post-processing, Plaques segmentation, Pyramid dilation convolution

# 1. Introduction

Cerebrovascular diseases are common diseases that seriously threaten the health of human beings [1]. Specially, stroke events [2,3] are a typical cerebrovascular disease, and also the second global leading cause of death, which kills millions of people. Moreover, stroke has two types of hemorrhagic stroke [4] and ischemic stroke [5], the latter accounts for the majority of all strokes [5,6]. Carotid arteries plaque is closely associated with ischemic stroke [7,8], and atherosclerotic in the carotid arteries is the reason of plaques formation [9], which is chronic and progressive process in the intimal layer of carotid artery walls. Furthermore, the plaques can decrease the blood vessel wall elasticity and reduce blood flow [10] to the brain. When plaques rupture, they may occlude a vessel in the brain leading to a stroke [11], thus, monitoring atherosclerotic is essential for the reduction of stroke incidents.

The stenosis degree of the carotid arteries is a reliable plaque measurement mode for assessing stroke risk [12] compared to intima-media thickness (IMT), which is not a strong cardiovascular event predictor [13], furthermore, total plaque area(TPA) is a stronger predictor for stroke [14]. Additionally, a plaque with moderate stenosis may induce distal embolization from the debris of the ruptured plaque [8,9]. This type of plaque called unstable plaque [15] has greater danger for patients, which has been widely studied in pathogenesis [16,17], statistical analysis[18] and imaging modalities [19].

Generally, three imaging modalities consisting of Magnetic resonance imaging (MRI) [20], computed tomography (CT) [21] and Ultrasound imaging have been used for carotid artery examination. Although MRI and CT can support high-resolution multi-slices for a plaque, MRI is expensive and takes a long time for scanning. CT has radiation, which is not suitable for some subjects. Whereas, ultrasound imaging has been more widely used for the diagnosis of atherosclerosis [10, 12] for its low-cost and non-radiation, however, this technique has its weakness such as low-resolution, artifact, and speckle noise, besides, plaque imaging quality depends on the ability of ultrasound technician.

Plaque features from ultrasound images, such as the shape, area, thickness, and components, are essential for a quantitative measurement of plaques. The plaques segmentation helps separate the diagnostic interest region from the background, and the texture features of the plaque are subsequently evaluated for its vulnerability. Manual segmentation by sonographers can be time-consuming and unrepeatable[14]. Measurement of TPA requires the segmentation of plaque boundaries and summing the plaque areas (PA) within the boundaries.

Classical snake[9,22] or level set[23,24], edge detector[25], hough transform[26] are used for plaques, or boundaries segmentation in carotid ultrasound images, but most of these methods depend on the manual features initialization and tedious procedures, may be negatively affected by local minima in noisy images. The automated feature extracted methods by CNN [8,14,27-29] have been extensively used for plaques or IMC segmentation, however, relative large-scale labeled datasets and region of interest (ROI) are required to train these models. Few-shot segmentation learning [30-33] can segment object regions using only a few annotated images, but the feature differences between the support and query images, and feature similarity between the target and background can weaken the predicted results, but there are few papers for plaque segmentation.

However, the mild plaque segmentation and IMT segmentation tasks are relatively easy compared with the vulnerable plaques with severe stenosis of our dataset. The difficulty can be summarized as three-fold in Figure 1. (1) obscure boundaries, caused by speckles and artifacts. (2) inconsistency between plaques. (3) plaque missing because of echoless effect.

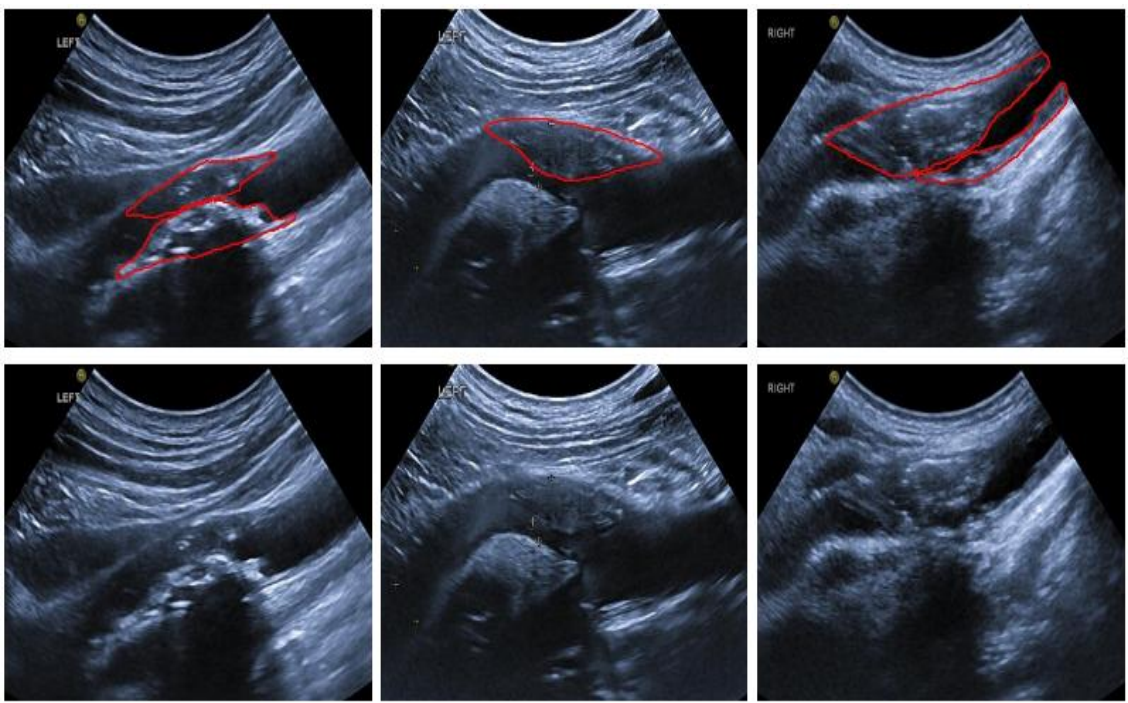

Fig. 1 Three ultrasound vulnerable plaque images, plaques are outlined in red in the first row.

In this study, we address the problem using a CNN-based approach without ROI pre-processing to automatically segment unstable plaques. We built a PA-Net inspired by FCN8 [34], the encoder of PA-Net has five scale semantic layers, the third to fifth layer of the encoder are used for parallel decoder using a low-stride. For beginning of each decoder, the pyramid dilation convolution (PDC), which is improved from the ASPP [35], is used for enlarging receptive field with small dilation rate to exploit available discrimination information. The three sub-networks are merged at the final decoder layer, because the fusion layers may have unconsidered features, we use the plug-and-play module SENet[36] for the channel attention. Moreover, the unstable plaques are located at the blood vessel lumens inside with no more than two separate areas, thus, we propose an efficient post-processing method to find the max contour or two max contours for refining the segmentation result.

automated feature extracted methods by CNN [8,14,27-29] have been extensively used for plaques or IMC segmentation, however, relative large-scale labeled datasets and region of interest (ROI) are required to train these models. Few-shot segmentation learning [30-33] can segment object regions using only a few annotated images, but the feature differences between the support and query images, and feature similarity between the target and background can weaken the predicted results, but there are few papers for plaque segmentation.

However, the mild plaque segmentation and IMT segmentation tasks are relatively easy compared with the vulnerable plaques with severe stenosis of our dataset. The difficulty can be summarized as three-fold in Figure 1. (1) obscure boundaries, caused by speckles and artifacts. (2) inconsistency between plaques. (3) plaque missing because of echoless effect.

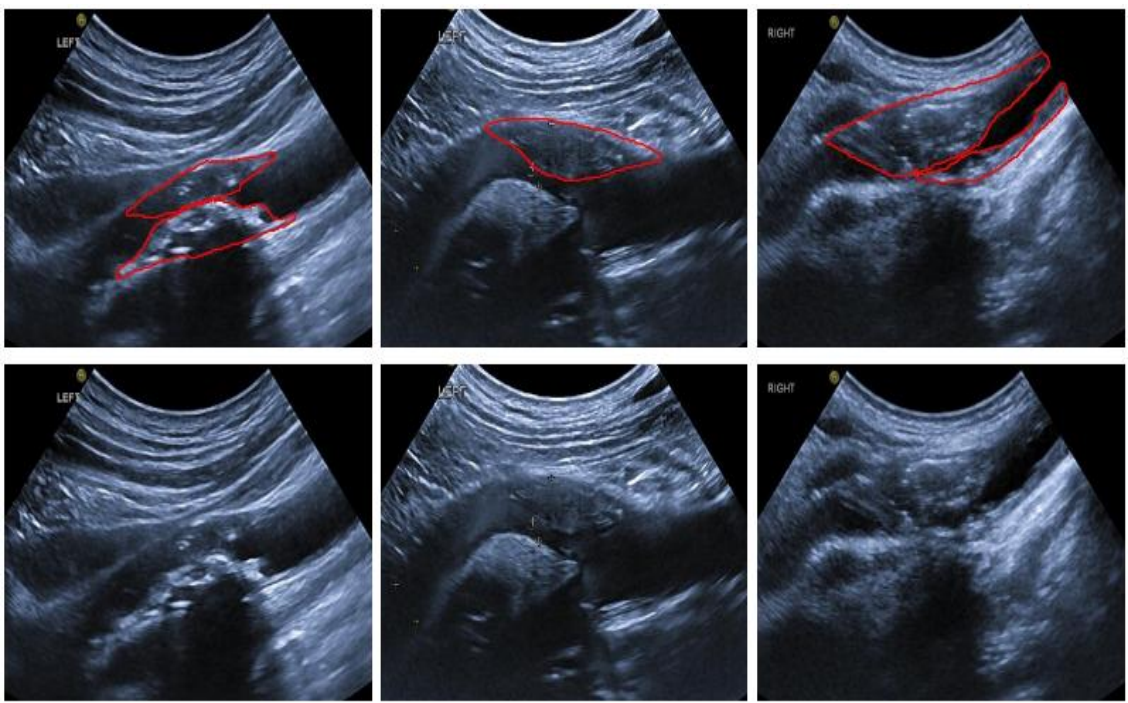

Fig. 1 Three ultrasound vulnerable plaque images, plaques are outlined in red in the first row.

In this study, we address the problem using a CNN-based approach without ROI pre-processing to automatically segment unstable plaques. We built a PA-Net inspired by FCN8 [34], the encoder of PA-Net has five scale semantic layers, the third to fifth layer of the encoder are used for parallel decoder using a low-stride. For beginning of each decoder, the pyramid dilation convolution (PDC), which is improved from the ASPP [35], is used for enlarging receptive field with small dilation rate to exploit available discrimination information. The three sub-networks are merged at the final decoder layer, because the fusion layers may have unconsidered features, we use the plug-and-play module SENet[36] for the channel attention. Moreover, the unstable plaques are located at the blood vessel lumens inside with no more than two separate areas, thus, we propose an efficient post-processing method to find the max contour or two max contours for refining the segmentation result.

Our contributions are as follows:

(A): A parallel decoder network with pyramid dilation convolutions, namely PA-Net, that are adjusted by the SENet module is proposed for segmentation.

(B): We propose a novel post-morphology method, which finds the max or second contours of the plaques to refine the result.

The rest of this paper is organized as follows: Section 2 first introduces our dataset, then the PA-Net structure, and morphology post-processing method, finally the implementation details. Section 3 presents our results for the plaques and some comparisons with the state-of-the-art methods. Further, we apply ablation experiments to our method. Section 4 presents the discussions. Finally, we conclude the paper in Section 5.

## 2. Materials and Methods

### 2.1 Dataset

The data used in this study were obtained from the Vessel Ultrasound Diagnostic Department of Xuan Wu Hospital, Capital Medical University. 30 subjects (with a mean age of 60.6 years) diagnosed with carotid stenosis were enrolled in our experiments in 2019, with over 60% confirmed from Doppler flow velocity measurements. The original data are longitudinally DICOM type with a resolution of $1024 \times 768$, acquired using the Hitachi ultrasonic machine (Japan).

The labeling was done by two experienced doctors using the ITK-SNAP[37] software, the adventitia contours of the internal carotid artery walls were drawn as a closed curve, which contains the plaque. Furthermore, the plaques outlines are extracted, and we fill them by OpenCV algorithm to finish the plaques masks.

### 2.2 Methods

The proposed method consists of a PA-Net with morphology post processing, as shown in Figure 2. The segmentation task comprises two stages: a training stage and a testing stage. In the training stage, the augmented images are inputted to the PA-Net for network parameter training from scratch. During the test stage, we apply the entire image to the Step-PDC net for segmenting the initial plaques; the image is then refined using the max profile post-processing method.

#### 2.2.1   PA-Net

Figure 2 shows the structure of our network, which is designed for end-to-end pixel-level segmentation. We have five-scale layers consisting of a series of convolutions and pooling operations to mine multi-scale semantic features. Specially, the $3 \times 3$ kernel convolutions, batch normalization, and rectified linear unit (ReLU) activation are combined for feature extraction. Besides, a 2D max pooling is used for reducing the dimension and redundant information.

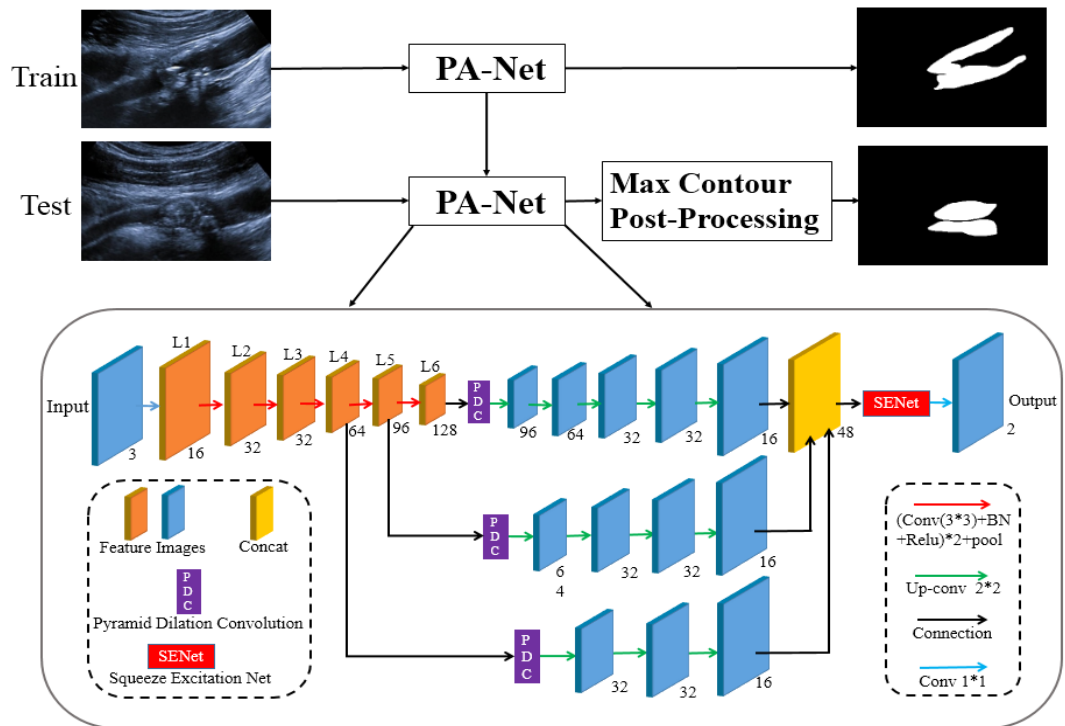

Fig. 2 PA-Net for segmenting carotid artery plaques in ultrasound images

The proposed method modifies the FCN8[34], the difference is the three skip concatenations, which are finished in the final layers. The three parallel decoding operations (from layers 4,5, and 6) are performed with up-sampling sizes of 32, 16, and 8. In addition, the up-sampling is done with a stride of 2 successively in each sub-network. In a word, the three sub-network confusion can explore multi-scale contextual information for discriminating complex plaque areas.

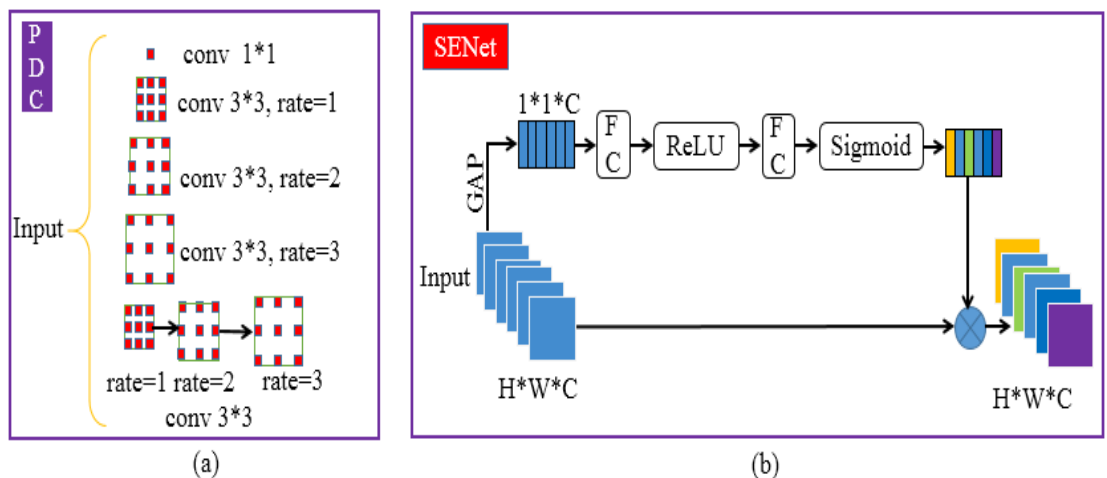

Fig. 3 (a) Pyramid dilation convolution and (b) Squeeze excitation net

Moreover, before the decoder of the three sub-networks, we add pyramid dilated convolutions[35,38,39] (PDC) with small dilated rates of 1, 2, and 3 to exploit more discriminative features on high-scale semantic layers as shown in Figure 3 (a). The dilated convolutions support expanding receptive fields without introducing more trainable parameters, and the small dilation factor can help better search for the local structures of the adjacent plaque.

The three sub-networks are simply merged at the end of the network, however, these fusion layers may comprise different semantic information, some of which contribute more while some contribute little, it is unsuitable for weighting all the features maps of the same weights. Notably, the squeeze excitation net (SENet)[36] can seek the relationship between the channels of the feature maps. Thus, the SENet module is applied after the fusion for features rectification, the key channels for the plaques discrimination can be weighted while the useless ones are weakened. Finally, we call our network PA-Net.

**2.2.2 Max Contour Morphology Post-processing**

The segmentation results obtained using the proposed network may contain some small non-plaque noises. Because plaques are aggregated in the near-wall or far-wall of the blood vessel, the unstable plaque is usually processed as a complete large plaque. Moreover, they have a high possibility to evolve into two individual plaques with similar areas, which are in the near and far walls of the arteries separately. Notably, these two plaques have no common regions in this situation.

Thus, we use this plaque prior feature for segmentation refinement. The post-processing pseudo-code for refining the segmented plaque can be described in Table 1.

Table 1 Post-processing method

| |
| --- |
| Input: initial segmentation |
| Output: final segmentation |
| Step 1: image erosion removing the small connection sections of the segmentation results, find all contours to get the contours number **NUM**. |
| Step 2: if **NUM** >1; find the max contour **C1** and second max contour **C2** of the segmentation. |
| Step 3:     if area(**N1**) < =5*area(**N2**); retain the two **C1 and C2** contours. |
| Step 4:     else; retain the max **C1** contour. |
| Step 5: else if **NUM** ==1; retain the only **C1** contour. |
| Step 6: fill the contours to finish the plaques. |

**2.2.3 Loss function**

The loss function is a metric to evaluate the error between the prediction and the ground truth. Some loss functions, such as cross-entropy loss function, and DICE loss function, SSIM loss function[40] have been used for semantic segmentation.

The DSC is defined in equation (1), L and P denote the label and predict, respectively.

$$\text{DICE loss} = 1 - \frac{2 * (L * P)}{L + P} \qquad (1)$$

The cross-entropy error loss is expressed as follows,

$$\text{crossentropy} = -[L * \log(P) + (1 - L) * \log(1 - P)] \tag{2}$$

The SSIM loss function is expressed in equation (3) and (4), specially, SSIM[41] is an image quality assessment metric based on the degradation of structural information. Moreover, the SSIM considers texture, contrast and luminance of an image which may be suitable for segmentation task.

$$\text{SSIM loss} = 1 - \text{SSIM}(L, P) \tag{3}$$

$$\text{SSIM}(L, P) = \frac{(2\mu_L \mu_P + C_1)}{(\mu_L^2 + \mu_P^2 + C_1)} * \frac{(2\sigma_{LP} + C_2)}{(\sigma_L^2 + \sigma_P^2 + C_2)} \tag{4}$$

Where $C_1$ and $C_2$ are constant, $\mu_L$ and $\mu_P$ are the mean values of L and P. $\sigma_L$ and $\sigma_P$ are the standard deviations of L and P.

## 2.3 Metric

The metrics used in this study for assessing the segmentation results of the plaques are the Dice (5), IoU (6), Acc (7), and modified Hausdorff distance (MHD)[42] (8). The Dice, IoU, and Acc are between 0 and 1, the greater the value, the better.

$$D\text{ice} = \frac{2|L \cap S|}{|L| + |S|} \tag{5}$$

Where L and S denote the label and segmentation by algorithm, respectively.

$$I\text{ou} = \frac{|L \cap S|}{|L| + |S| - |L \cap S|} \tag{6}$$

$$\text{Acc} = \frac{TP + TN}{TP + TN + FP + FN} \tag{7}$$

Where TP, TN, FP, and FN are the true positive, true negative, false positive, and false negative, respectively.

$$\text{MHD} = \max[d(L,S), d(S,L)] \tag{8}$$

$$\text{Where} \quad d(L,S) = \frac{\sum_{a \in L} d(a,S)}{N_a} \tag{9}$$

The MHD determines the mean distance between the predict and the label, the lower the value, the better the result. The lowest value is 0, which indicates that two images are the same.

## 2.4 Implementation details

We used a Tesla V100 GPU with 16G memory for data training using the Keras framework and TensorFlow as the backbone. A 10-fold cross validation was adopted because only 30 images were used, specially, 3 for testing, and 27 for training. The one-hot code was adopted for the pixel softmax classification.

We performed the following data augmentation operations to increase samples number to 180: horizontal transformation, vertical transformation, and 180 ° flip, +30 ° and −30° affine rotations, and elastic deformation. The Adam optimizer was chosen for the optimization; the learning rate was set to 0.001, and the learning epoch was 100.

## 3. Results

### 3.1. Comparison with some cutting-edge methods

Carotid artery unstable ultrasound images segmentation experiments are performed by our method and some methods (U-Net[43], SegNet[44], FCN8[34], Attention U-Net[45], DeepLabv3[35], GCN[46]). It is worth noting that all the images are refined by the max-contour post-processing method, Table 2 lists the results of nine different methods. Figure 7 shows five images of the carotid artery ultrasound along with the label and the different methods used. Moreover, these methods were provided with the same augmentation data, the same loss function, and training method to ensure a fair comparison.

As can be seen from Table 2, our proposed method yields the best result in terms of the Dice (0.820±0.066), IoU (0.701±0.094), Acc (0.969±0.014), and MHD (1.43±1.27) compared to the other methods. In addition, U-Net, SegNet, FCN8, Attention U-Net, DeepLabv3, and GCN obtain the dice metrics 0.808±0.071, 0.747±0.088, 0.803±0.073, 0.796±0.079, 0.798±0.070, 0.784±0.078, respectively. And our method exceeds them on dice metric by 0.012, 0.073, 0.017, 0.024, 0.022, 0.036. Furthermore, they acquire MHD metrics 1.92±1.94, 2.99±2.11, 1.72±1.54, 1.98±1.79, 2.03±1.60, and 2.36±2.24, the proposed method decreases by 0.49, 1.56, 0.29, 0.55, 0.60, 0.93 on the MHD metric.

Table 2 Comparison between different methods after post-processing, where the numbers in bold indicate the best result, (Mean ±SD)

| Metric | Dice | IoU | Acc | MHD |
|---|---|---|---|---|
| Proposed | **0.820±0.066** | **0.701±0.094** | **0.969±0.014** | **1.43±1.27** |
| U-Net | 0.808±0.071 | 0.684±0.099 | 0.965±0.015 | 1.92±1.94 |
| SegNet | 0.747±0.088 | 0.603±0.109 | 0.953±0.019 | 2.99±2.11 |
| FCN8 | 0.803±0.073 | 0.677±0.100 | 0.965±0.016 | 1.72±1.54 |
| Attention U-Net | 0.796±0.079 | 0.669±0.107 | 0.963±0.021 | 1.98±1.79 |
| DeepLabv3 | 0.798±0.070 | 0.670±0.097 | 0.964±0.019 | 2.03±1.60 |
| GCN | 0.784±0.078 | 0.651±0.101 | 0.960±0.017 | 2.36±2.24 |

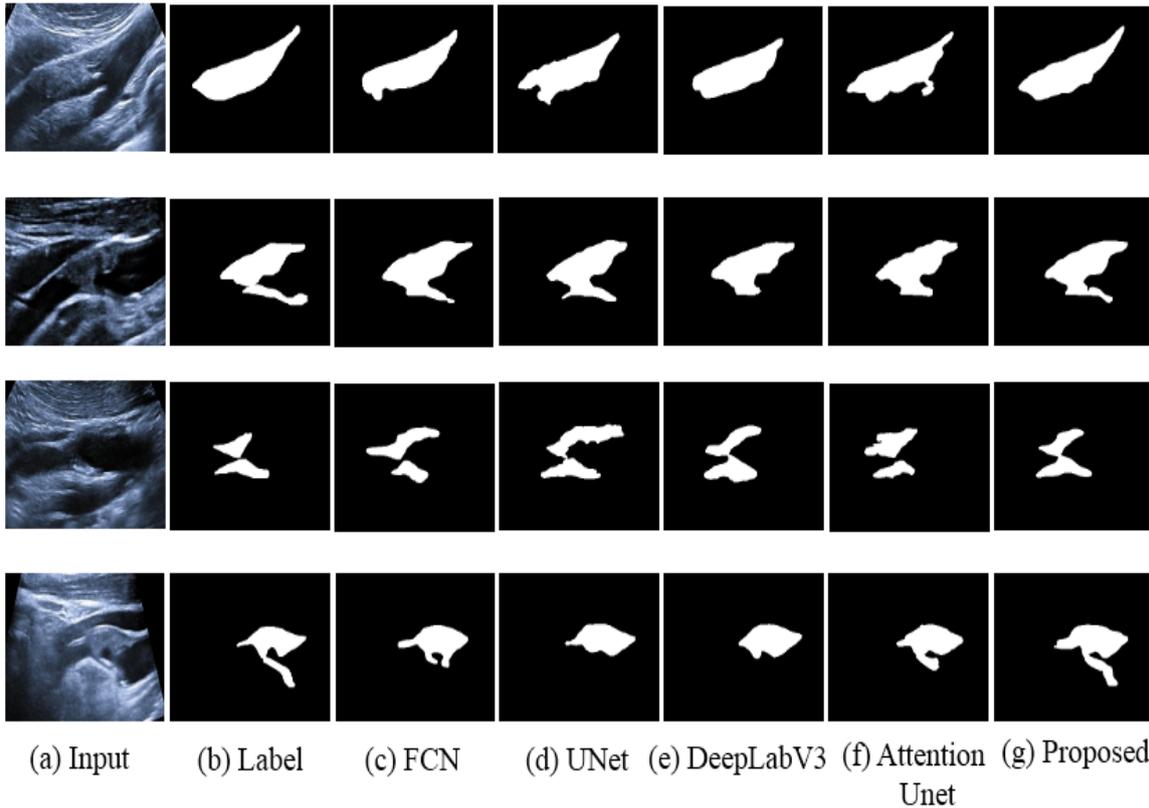

Fig. 4 Results obtained using different methods: (a) Original images (b), Labels, (c) FCN (d), U-Net, (e) DeepLabV3, (f) Attention U-Net, (g) Proposed

Figure 4 shows the results of four different images from the data obtained using the seven methods listed in Table 2, our method outperforms the others in terms of the visual result. As can be seen from the first and second rows of Figure 4, our method obtain more precious results compared to other methods, the results of which are closer to the labels. Besides, our method is more robust to the false positive and blurry boundaries near the plaque and can discriminate the low-contrast plaque from the noisy images in row 3 of Figure 4. Moreover, the proposed method acquires more intact the plaque contour from the confused boundary, while the results of other methods are coarse, and may discard plaque regions. In brief, our method provides reasonable results in terms of the total shapes.

The results before post-processing of the seven methods are shown in Table3, our method also obtain the bests compared to other methods. Figure 5 shows the Dice results of the seven methods before and after the morphological processing. After post-processing, the Dice metrics of the proposed, U-Net, SegNet, FCN8, Attention U-Net, DeepLabv3, GCN increase by 0.008,0.005,0.015,0.008,0.011,0.005,0.006, respectively. For all the methods, the post-processing improves the final result, the average increase is 0.0081. Thus, the proposed post-processing demonstrates a good process capacity for fine segmentation.

Table 3 Comparison between different methods before post-processing, where the numbers in bold indicate the best result, (Mean ±SD)

| Metric | Dice | IoU | Acc | MHD |
| --- | --- | --- | --- | --- |
| Proposed | **0.812±0.068** | **0.690±0.095** | **0.969±0.014** | **1.71±1.37** |
| U-Net | 0.803±0.067 | 0.676±0.091 | 0.966±0.014 | 2.17±2.04 |
| SegNet | 0.732±0.089 | 0.586±0.108 | 0.952±0.019 | 3.17±1.93 |
| FCN8 | 0.795±0.068 | 0.665±0.090 | 0.964±0.014 | 2.02±1.47 |
| Attention U-Net | 0.785 ±0.080 | 0.653 ±0.103 | 0.962±0.020 | 2.67±2.51 |
| DeepLabv3 | 0.793±0.069 | 0.662±0.092 | 0.963±0.018 | 2.18±1.70 |
| GCN | 0.778±0.079 | 0.643±0.100 | 0.960±0.017 | 2.89±2.61 |

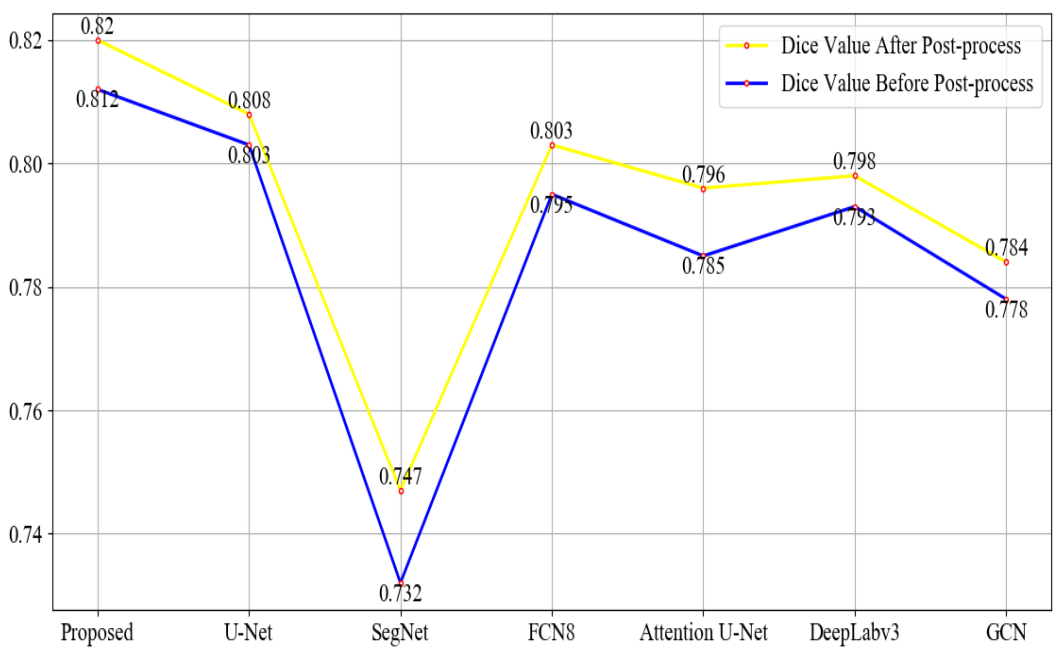

Fig. 5 Post-processing effects on the seven methods

Figure 6 shows the three representative cases with respect to the proposed max contour morphology process method, which depends on the nature of the plaques. In particular, the post-processing technique first acquires all the contours in the initial segmentation, then calculates the areas of all contours, finally, the small non-plaque parts out of the blood vessels of the initial segmented plaques are eliminated using the threshold values set in 2.2.2 section. Thus, the post-processing method is capable of realizing the coarse-to-fine segmentation performance.

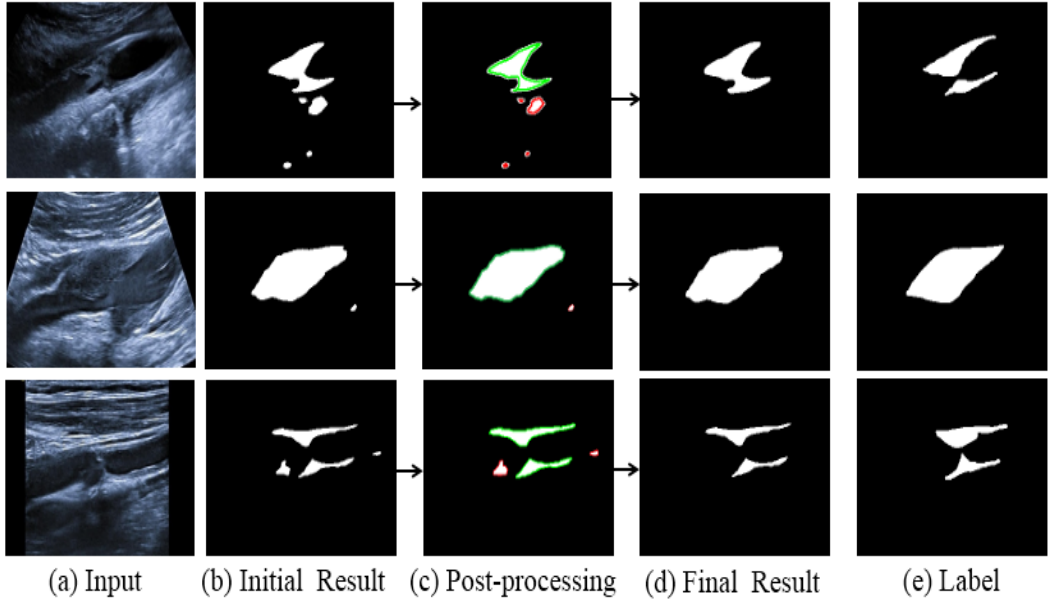

Fig. 6 Post-processing of three cases of plaques

## 3.2. Ablation experiments

To validate the validity of each section of the proposed method, we applied ablation experiments. Firstly, we experiment the three loss functions for the plaques segmentation evaluation. Second, the proposed modules are verified respectively.

**Loss Functions:** Table 4 lists the results of three loss functions, the other parts of the proposed method remain the same. As can found that SSIM loss function obtains the results on the four metrics Dice, IOU, ACC, MHD, of 0.801±0.058, 0.671±0.079, 0.966±0.013, 1.89±1.45, respectively. Binary cross entropy loss function obtains results of the four metrics 0.807±0.064, 0.682±0.091, 0.967±0.016, 1.83±1.51. Dice loss function obtains the results of these metrics 0.812±0.068, 0.690±0.095, 0.969±0.014, 1.71±1.37. The dice loss function acquires the best plaques segmentation results on the four metrics compare to SSIM loss function and Binary cross entropy loss function.

**Table 4** Performance of loss functions (Mean ±SD)

| SSIM loss | Binary Cross loss | Dice loss | Dice | IOU | ACC | MHD |
|---|---|---|---|---|---|---|
| √ | × | × | 0.801±0.058 | 0.671±0.079 | 0.966±0.013 | 1.89±1.45 |
| × | √ | × | 0.807±0.064 | 0.682±0.091 | 0.967±0.016 | 1.83±1.51 |
| × | × | √ | **0.812±0.068** | **0.690±0.095** | **0.969±0.014** | **1.71±1.37** |

**Ablated Experiments:** We ablate the sub-methods of the proposed method and the results are as listed in Table 5. The parallel networks are the base networks, which are used in each ablated method. The base network obtains the results on Dice metric, IOU metric, ACC metric, MHD metric of 0.798±0.066, 0.669±0.090, 0.966±0.014, 2.12±1.74. With the introducing module of PDC, SE-Net, and Post-processing separately or combinedly, the segmentation results all increase to a certain degree. These shows the validity of each proposed module.

Table 5 Ablation of the sub-methods of each section (Mean ±SD)

| B-N | PDC | S-N | P-P | Dice | IOU | Acc | MHD |
|---|---|---|---|---|---|---|---|
| √ | × | × | × | 0.798±0.066 | 0.669±0.090 | 0.966±0.014 | 2.12±1.74 |
| √ | √ | × | × | 0.805±0.063 | 0.678±0.088 | 0.967±0.015 | 1.66±1.21 |
| √ | × | √ | × | 0.805 ±0.065 | 0.679 ±0.090 | 0.966±0.018 | 1.80±1.27 |
| √ | √ | √ | × | 0.812±0.068 | 0.690±0.095 | 0.969±0.014 | 1.71±1.37 |
| √ | √ | √ | √ | **0.820±0.066** | **0.701±0.094** | **0.969±0.014** | **1.43±1.27** |

## 4. Discussion

Carotid arteries unstable plaque is a reliable factor for screening atherosclerosis, early detecting unstable plaques can help to take the necessary measures to reduce the occurrence of stroke. However, the unstable plaques may be contaminated by various noise, and they have no fixed shape in terms of their appearance, in addition, there is uncertainty for the plaques segmentation that are without a prior knowledge unlike in the natural image segmentation. Consequently, these factors increase the difficulties for the plaques segmentation in carotid ultrasound images.

In this paper, we propose a novel CNN for the small dataset, with its base net originating from FCN8. The results of the base net and FCN8 are similar, however, our base net result is slightly higher than that of FCN8. When we add the PDC for more reception fields and the squeeze excitation net for the weight adjustment of the different channels, the segmentation result becomes better. Additionally, the dice loss function is more suitable for the segmentation task. Compare to some state-of-the-art methods, our method outperforms them on the four metric, furthermore, the post-processing module take effects on all the methods for the segmentation result refining. In other words, each proposed module makes the result better.

The shortcoming of our method is that the 30 images dataset used in this study is insufficient to represent the actual plaque distribution condition, and the proposed network may not generalize well for new data. Apart from that, the post-processing should be further improved because it may abandon plaque regions. In fact, more data can decrease the over-fitting during training and improve the generalization for the neural network. In the future, more data from multi-centers hospitals should be obtained, and our method should also be modified to improve the segmentation result.

## 5. Conclusion

In this study, we proposed a novel automatic CNN method for unstable plaque identification. We improved the FCN8 decoding section, named as PA-Net, a PDC module was proposed for enlarging the reception field in the three sub-networks and a squeeze excitation network for readjusting the weights of the final cascaded channels for the channel attention. Furthermore, a max contour area morphological post-processing method was applied to further refine the plaque result. Compare to SSIM loss function and binary cross entropy loss function, dice loss function is better for plaques segmentation. The proposed method outperformed previous methods in terms of the Dice and MHD results. Moreover, the ablation experiment shows that each section of our method is essential.

The proposed automatic plaque segmentation method can be useful for general clinicians to quantify the morphological features of vulnerable plaques, and to improve the objectivity and efficiency of plaque interpretation as an auxiliary method in clinic.

## Declaration of Competing Interest

There are no conflicts of interest.

### ACKNOWLEDGEMENTS

This work is supported by Beijing Natural Science Foundation (Grant Number: Z200024), the National Key Research and Development Program of China (Grant Number: 2016YFF0201002), the University Synergy Innovation Program of Anhui Province (Grant Number: GXXT-2019-044).